\documentclass[12pt]{iopart}

\usepackage{graphicx}
\usepackage{bmpsize}
\usepackage{dcolumn}
\usepackage{bm}
\usepackage[caption=false]{subfig}
\usepackage{iopams}
\usepackage{float}
\usepackage{lineno}
\usepackage{cite}

\renewcommand\thesubsection{\Arabic{subsection}}

\begin{document}
\title{Experimental Realization of a Relativistic Harmonic Oscillator}

\author{Kurt M. Fujiwara$^1$\footnote{Equal contributions.}, Zachary A. Geiger$^1$$\ddagger$, Kevin Singh$^1$, Ruwan Senaratne$^1$, Shankari V. Rajagopal$^1$, Mikhail Lipatov$^1$, Toshihiko Shimasaki$^1$, David M. Weld$^1$}
 \ead{weld@ucsb.edu}
\address{$^1$University of California and California Institute for Quantum Emulation, Santa Barbara CA 93105}

\begin{abstract}
We report the experimental study of a  harmonic oscillator in the relativistic regime. The oscillator is composed of Bose-condensed lithium atoms in the third band of an optical lattice, which have an energy-momentum relation nearly identical to that of a massive relativistic particle, with an effective mass reduced below the bare value and a greatly reduced effective speed of light.   Imaging the shape of oscillator trajectories at velocities up to 98\% of the effective speed of light reveals a crossover from sinusoidal to nearly photon-like propagation. The existence of a maximum velocity causes the measured period of oscillations to increase with energy; our measurements reveal beyond-leading-order contributions to this relativistic anharmonicity. We observe an intrinsic relativistic dephasing of oscillator ensembles, and a monopole oscillation with exactly the opposite phase of that predicted for non-relativistic harmonic motion. All observed dynamics are in quantitative agreement with longstanding but hitherto-untested relativistic predictions.

\end{abstract}

\maketitle

The harmonic oscillator has been a concept of central significance in physics and technology since Galileo's observations of constant-period pendulum motion. The advent of special relativity suggested a simple question: what happens to this archetypal physical system when the maximum oscillator velocity approaches the speed of light? While theories of the relativistic harmonic oscillator have been discussed for decades~\cite{RHO-nature,RHO_amjp,RHO-mathphys,RHO_PRE,Parker_gravtrain}, the combination of special relativity and harmonic motion has proven resistant to physical realization. The infeasibility of realizing harmonic traps with depths on the order of a particle's rest mass energy (Boltzmann's constant times nearly six billion degrees Kelvin for an electron) has motivated work studying relativistic phenomena in disparate physical contexts, including a measurement of the Dirac oscillator spectrum in an array of microwave resonators~\cite{RHO-microwave} and proposals and realizations of effective relativistic effects in trapped atoms~\cite{juzeliunas-relativistic,klein-weitz,zitter-spielman,zitter-engels,weitz-veselago,weitzrelativity,lewenstein-unruh}, trapped ions~\cite{delgado-dirac,zitter-blatt,klein-blatt}, photonic waveguides~\cite{Longhi-RHO}, and graphene~\cite{reviewkleintunnelinggraphene}. 

Here we report the experimental realization of a harmonic oscillator in the relativistic regime, using ultracold atoms moving in the third band of an optical lattice. Optical lattices are an ideal context in which to study dynamics in excited bands: previous experiments in higher bands have, for example, observed Bloch oscillations~\cite{salomon-blochPRA}, demonstrated coherent matter-wave behavior in higher-bands~\cite{hemmerich-pbandSF,hemmerich-orbitalreview}, studied issues relevant to quantum transport~\cite{quantumconveyor}, measured band nonlinearities and momentum-space Fermi-gas dynamics~\cite{sengstock-photoconductivity} and demonstrated precise wavepacket manipulation~\cite{sherson-pumpprobe,sherson-wavepackets}. The results reported here depend crucially upon the ability to image the time evolution of the position of atomic wavepackets in higher bands.

\begin{figure}[t!]
\begin{center}
\includegraphics[width=.65\linewidth]{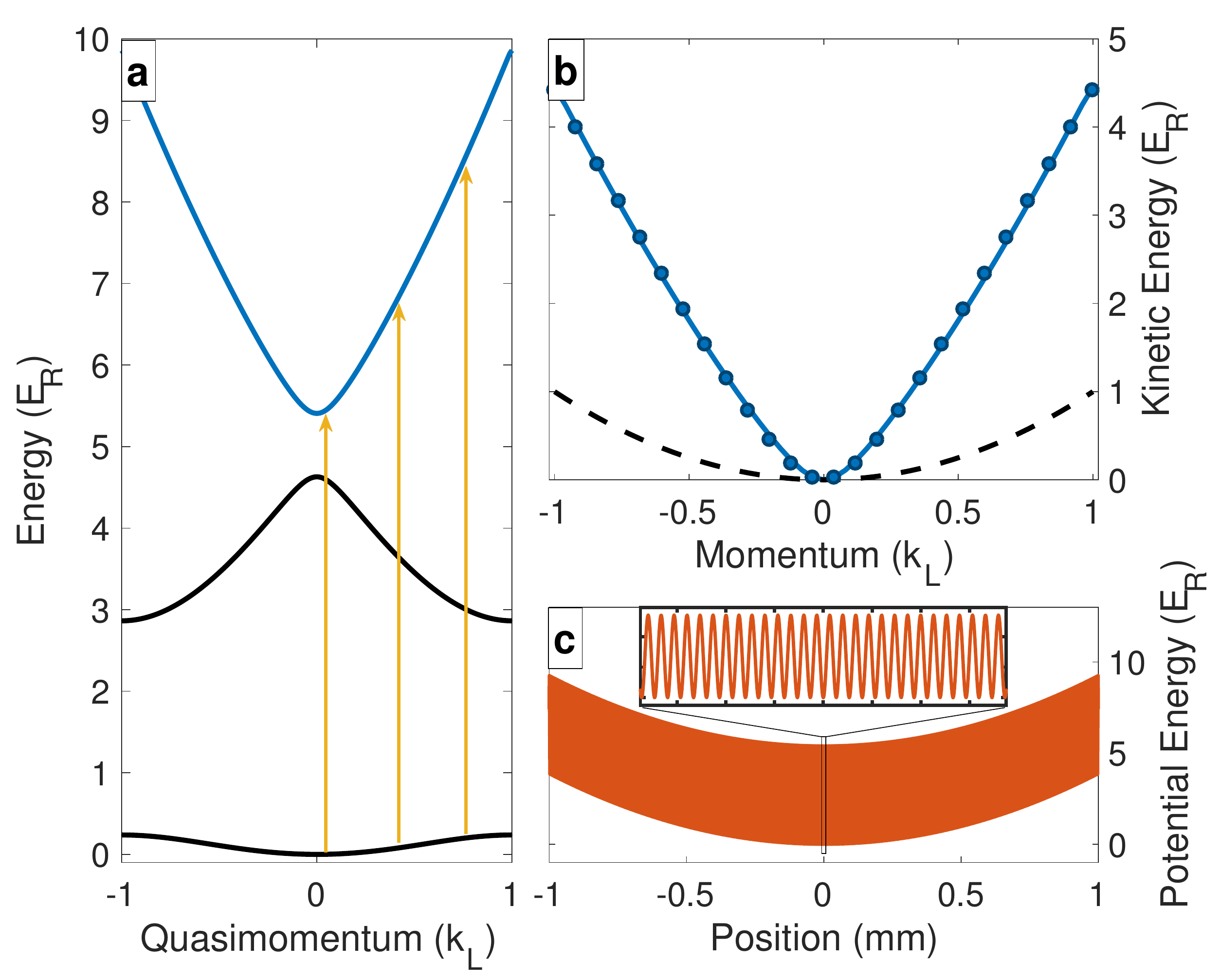}\label{fig:bands}
\end{center}
\caption{Realizing a relativistic harmonic oscillator. \textbf{a:}~Calculated band structure of 5.4-$E_\mathrm{R}$-deep optical lattice. Here $E_\mathrm{R}=\hbar^2 k_\mathrm{R}^2/2m_\mathrm{Li}$ is the recoil energy, $k_\mathrm{R}=2\pi/\lambda$ is the lattice wavevector, and $\lambda=1064$~nm is the wavelength of the lattice laser. Arrows indicate the 137, 170, and 210~kHz excitations used to prepare the three oscillators shown in figure~\ref{fig:odmix}. \textbf{b:}~Calculated dispersion relation for the third band (solid line), compared to the relativistic kinetic energy fit (circles). Note the close agreement throughout the Brillouin zone. The free particle dispersion (dashed line) is also shown. \textbf{c:}~Position-dependent potential energy due to the combination of the lattice and the harmonic confinement.  Inset shows a horizontally magnified view of the central 15~$\mu$m of the trap.} 
\label{fig:RHO_disp}
\end{figure}

Einstein's  relativistic dispersion relation 
\begin{equation}
\mathcal{E_\mathrm{rel}}=\sqrt{p^2c^2+m^2c^4}=\mathcal{E}_\mathrm{kin}+mc^2,
\label{reldisp}
\end{equation}
where $\mathcal{E_\mathrm{rel}}$ is the energy, $\mathcal{E}_\mathrm{kin}$ the kinetic energy, $p$ the momentum, $m$ the effective rest mass, and $c$ the effective speed of light, is achieved in our experiments by placing non-interacting Bose-condensed lithium atoms into the third band of a 1D optical lattice. While it is common to speak informally of any approximately linear higher-band dispersion relation as having ``relativistic'' character, here we demonstrate a near-exact quantitative correspondence, which holds throughout the Brillouin zone, between the third-band dispersion for the particular lattice parameters we choose and the relativistic kinetic energy of a massive particle (see figure \ref{fig:RHO_disp}b). Motion of atoms in the third band will thus have relativistic character to an excellent approximation: atomic trajectories will trace out the worldlines of relativistic particles with effective rest mass $m$ and an effective speed of light $c$. The addition of a harmonic potential allows the direct experimental study of the worldlines and dynamics of a relativistic harmonic oscillator, with total energy given by $\mathcal{E}=mc^2+\mathcal{E}_\mathrm{kin}+m\omega^2x^2/2,$ where $\omega$ is the harmonic oscillator frequency in the non-relativistic limit and $x$ is displacement.

Figure~\ref{fig:RHO_disp}b compares the third-band dispersion relation to a relativistic fit of the form of equation~\ref{reldisp}, with an effective mass $m$ and effective speed of light $c$ treated as fit parameters. Here, and throughout this article, $c=143\ $mm/s and $m=0.07\ m_\mathrm{Li}$ (roughly half a proton mass). $m_\mathrm{Li}$ represents the bare atomic mass of lithium in the absence of the lattice. Crucially, the relativistic fit shows excellent agreement throughout the Brillouin zone. The origin of this striking correspondence is the zero-quasimomentum avoided crossing between the second and third bands. For momenta much smaller or larger than a lattice recoil, modifications to the relativistic dispersion are anticipated; these momenta are outside the range experimentally probed in this work. It is important to note that the results we report explore the relativistic physics of normal matter, not antimatter, as our band structure does not model the negative-energy branch of the relativistic dispersion. In particular, our relativistic fit involves only the third band, and the gap between the second and third bands is not equal to twice the rest mass energy.

Before describing the experiments it is useful to briefly outline existing predictions for the behavior of a relativistic harmonic oscillator~\cite{RHO-nature,RHO_amjp,RHO-mathphys,RHO_PRE,Parker_gravtrain}. As the maximum velocity $v_\mathrm{max}$ approaches $c$,  relativistic effects are expected to modify the character of harmonic motion in several ways. The shape of the oscillator's trajectory changes from the familiar sinusoidal form in the non-relativistic limit to an increasingly triangular shape indicative of nearly photon-like propagation, with worldline curvature more and more concentrated at the turning points. Physically, this is because the increasingly linear dispersion makes the force due to the harmonic confinement less and less effective at changing the particles' velocity as it approaches the effective speed of light $c$. For the same reason, the oscillation amplitude diverges as $\beta_\mathrm{max}\equiv v_\mathrm{max}/c$ approaches 1. Relativistic anharmonicity is manifested by an increase in period with energy and a characteristic monopole oscillation and dephasing of oscillator ensembles initialized with a range of energies. All these behaviors can be quantitatively captured by an exact theoretical model~\cite{RHO_amjp}. The relation between coordinate displacement $x$ and coordinate time $t$ that constitutes the theoretically predicted trajectory or worldline of a relativistic oscillator in the lab frame is given by the equations
\begin{linenomath}\begin{eqnarray}
\omega t &=& \sqrt{2\left(\gamma_0+1\right)}E\left(\phi,\kappa\right) - \sqrt{\frac{2}{\gamma_0+1}}F\left(\phi,\kappa\right)\label{eq:worldlines1}
\\
\omega x/c &=& \sqrt{2(\gamma_0-1)}\sin\phi
\label{eq:worldlines2}
\end{eqnarray}\end{linenomath}
where $\gamma_0\equiv\mathcal{E}/mc^2$, the maximum Lorentz factor, is the total energy in units of the rest energy, $\kappa=\sqrt{(\gamma_0-1)/(\gamma_0+1)}$, and $F$ and $E$ are the incomplete elliptic integrals of the first and second kind respectively~\cite{RHO_amjp}.  Setting $\phi=2\pi$ in equation~\ref{eq:worldlines1} gives an exact expression for the period $T$ of the motion, which can be Taylor expanded to lowest order in $(\gamma_0-1)$ as~\cite{RHO-nature}
\begin{equation}
\frac{T}{T_0}\simeq 1+\frac{3}{8}(\gamma_0-1)=1+\frac{3}{16}\frac{\omega^2}{c^2}A^2,
\label{eqperiod}
\end{equation}
where $T_0$ is the oscillator period in the nonrelativistic limit, and the oscillation amplitude $A$ is related to the total energy as
$A=c\sqrt{2(\gamma_0-1)}/\omega$.  At sufficiently high energies, the trajectories described by equations~\ref{eq:worldlines1} and~\ref{eq:worldlines2} exhibit all the relativistic effects on harmonic motion which are qualitatively described above. Experimental observation and measurement of such trajectories is the main subject of this work.

\begin{figure}
\begin{center}
\includegraphics[width=0.6\linewidth]{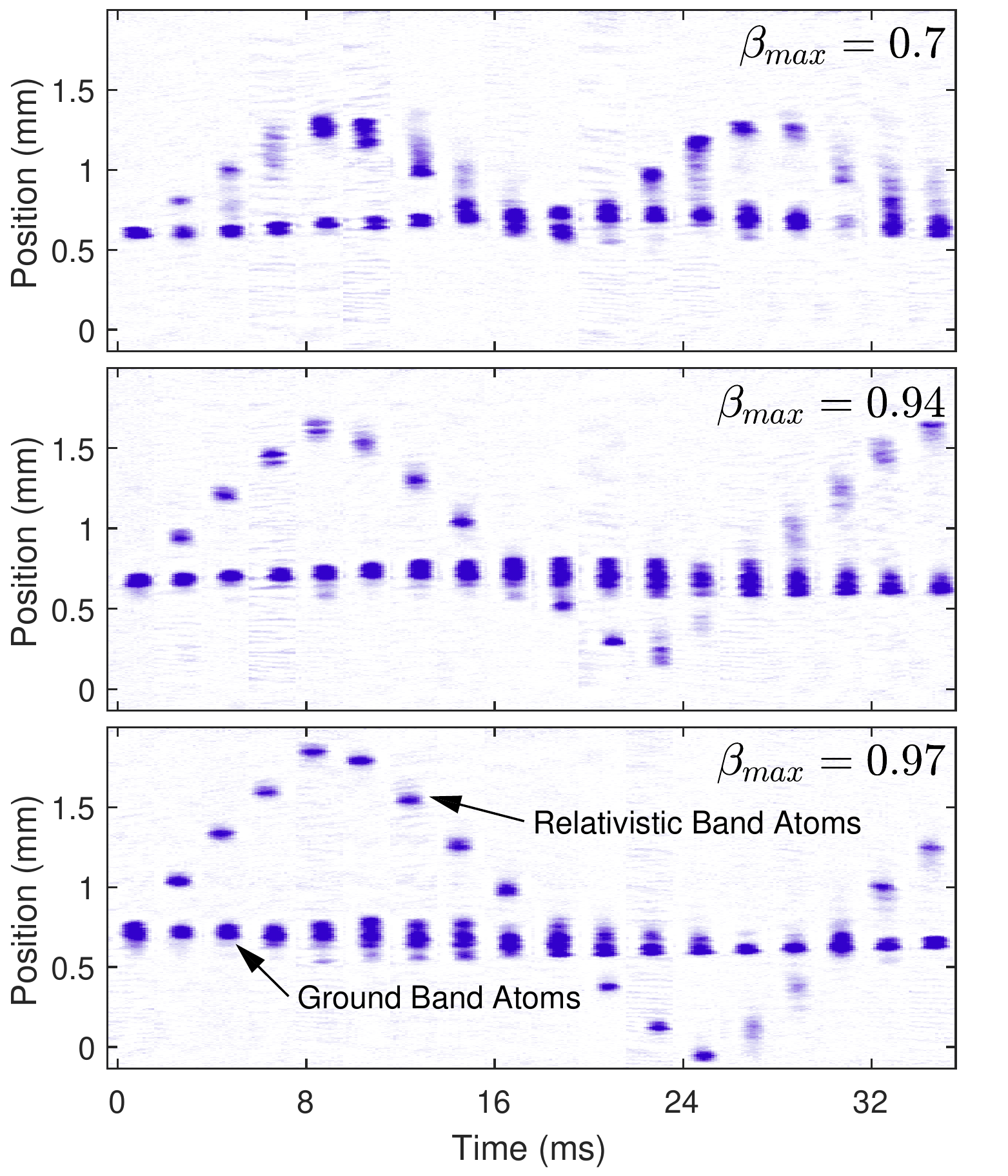}
\caption{Measured trajectories of relativistic harmonic oscillators.
Each panel shows a time sequence of in-situ absorption images of a non-interacting atomic cloud with population in the first and third bands of an optical lattice. Third-band atoms trace out relativistic harmonic oscillator worldlines of increasing amplitude and period. Ground-band atoms Bloch oscillate~\cite{RSBO-arxiv} around their original position, and are irrelevant to the relativistic dynamics. The maximum value of $\beta=v/c$ indicated in each panel was attained by loading the third band at varying quasimomenta, using excitation frequencies of (from top to bottom) 137, 170, and 210~kHz and corresponding hold times before excitation of 0~ms, 8~ms, and 16~ms.}
\label{fig:odmix}
\end{center}
\end{figure}

Here we briefly describe our procedure for Bose-condensing lithium and loading atoms into the third band with tunable energy. The low mass, which gives rise to rapid dynamics, and the shallow scattering length zero-crossing of the low-field Feshbach resonance~\cite{Hulet-tunableinteractionsin7Li} make Lithium particularly suitable for excited-band experiments.
The experiments are initiated by creating a Bose condensate of approximately 10$^5$~${}^7$Li atoms in the $|F=1,m_F=1\rangle$ state at a temperature of 20~nK, produced by evaporation in an optical dipole trap after loading a magneto-optical trap from a collimated atomic beam~\cite{nozzleRSI} and precooling with gray molasses~\cite{LiD1graymolasses,hamiltonGM} and RF evaporation. Atom-atom interactions are set to zero after evaporation by ramping an applied magnetic field to the $s$-wave scattering length zero-crossing of a Feshbach resonance, at approximately 543.6~G~\cite{Hulet-tunableinteractionsin7Li}. Atoms are then loaded into the ground band of a 1D optical lattice of depth 5.4~$E_\mathrm{R}$, produced by a retroreflected laser of wavelength $\lambda=1064$~nm focused to a  150~$\mu$m beam waist.  Lattice beam power is ramped to the final depth in 100~ms, keeping the optical trap power constant. Lattice depth is measured by a combination of matter-wave diffraction from pulsed lattices~\cite{KapitzaDiracPritchard} and amplitude-modulation spectroscopy~\cite{BECinlattice-RolstonPhillips}. The lattice depth, laser wavelength, and bare mass of $^7$Li determine the effective mass $m$ and effective speed of light $c$ in equation~\ref{reldisp}.
Magnetic field curvature from external coils generate the overall harmonic confinement with frequency $\nu=16.6$~Hz, centered approximately 0.3~mm from the original BEC position along the lattice direction.  Note that this frequency is measured for bare ${}^7$Li atoms, so differs from the frequency $\omega/2\pi$ by the square root of the ratio between the bare and effective masses. 
When the optical trap beams are suddenly switched off, the atoms evolve in the combined potential of this harmonic trap and the optical lattice.
For atoms in the ground band of the lattice, the evolution in the combined potential takes the form of Bloch oscillations~\cite{Bloch1929,Zener1934,salomon-blochoscs,nagerlinteractionBOs,RSBO-arxiv}. Relativistic dynamics are initiated after a variable hold time by amplitude modulation of the lattice beam intensity which resonantly excites a fraction of the atoms to the third band. Because the quasimomentum depends on the phase of the Bloch oscillation and the excitation conserves quasimomentum, appropriately timed resonant modulation pulses can initialize the atomic ensemble with tunable initial kinetic energy as diagrammed in figure~\ref{fig:RHO_disp}. For the experiments reported here the pulse duration and depth are 500~$\mu$s and 20\% respectively.  This method is used to prepare relativistic-band harmonic oscillators with varying values of the total energy.

We investigate the dynamics of this relativistic harmonic oscillator by measuring the atomic spatial distribution using absorption imaging after variable hold time in the third band. Figure~\ref{fig:odmix} shows a sequence of such images which  map out the evolution for three different values of $\beta_\mathrm{max}$.  The relativistic anharmonicity is clearly evident: as the energy increases, the period and amplitude of the oscillation increase, and the trajectories become less sinusoidal. For velocities comparable to the effective speed of light, atomic trajectories are straight for the majority of the period, with curvature concentrated near the turning points, in agreement with expectations for a highly relativistic oscillator. The atoms which remain in the ground band are also visible in the images; they continue to Bloch oscillate near their original position during this time~\cite{RSBO-arxiv}. Due to the absence of interactions, ground-band atoms do not affect higher-band relativistic dynamics, though on a practical level spatial overlap with ground-band atoms can impede imaging of very low-energy (non-relativistic) higher-band oscillations.  

\begin{figure}[t!]
\begin{center}
\includegraphics[width=0.7\linewidth]{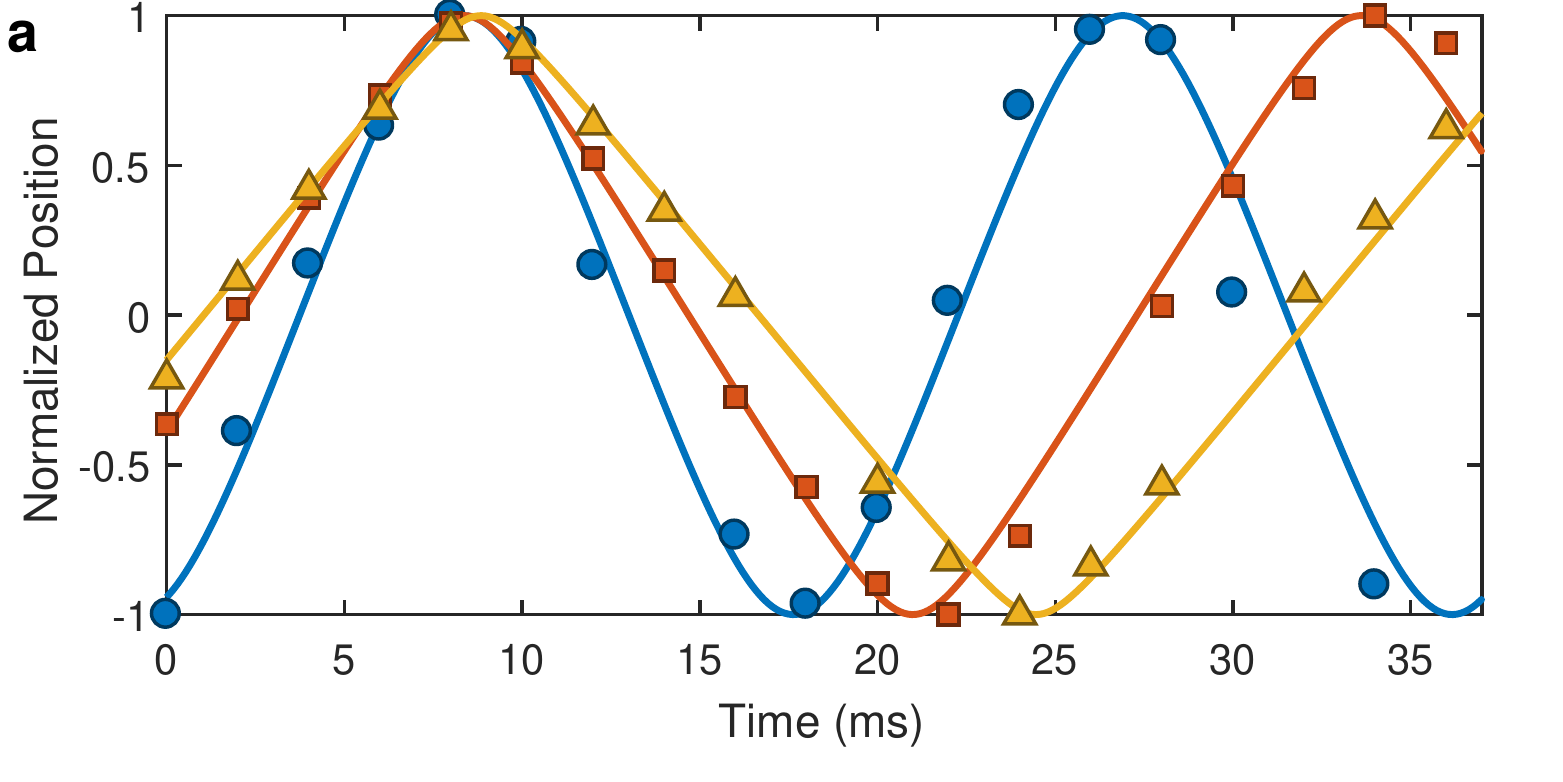}
\includegraphics[width=0.7\linewidth]{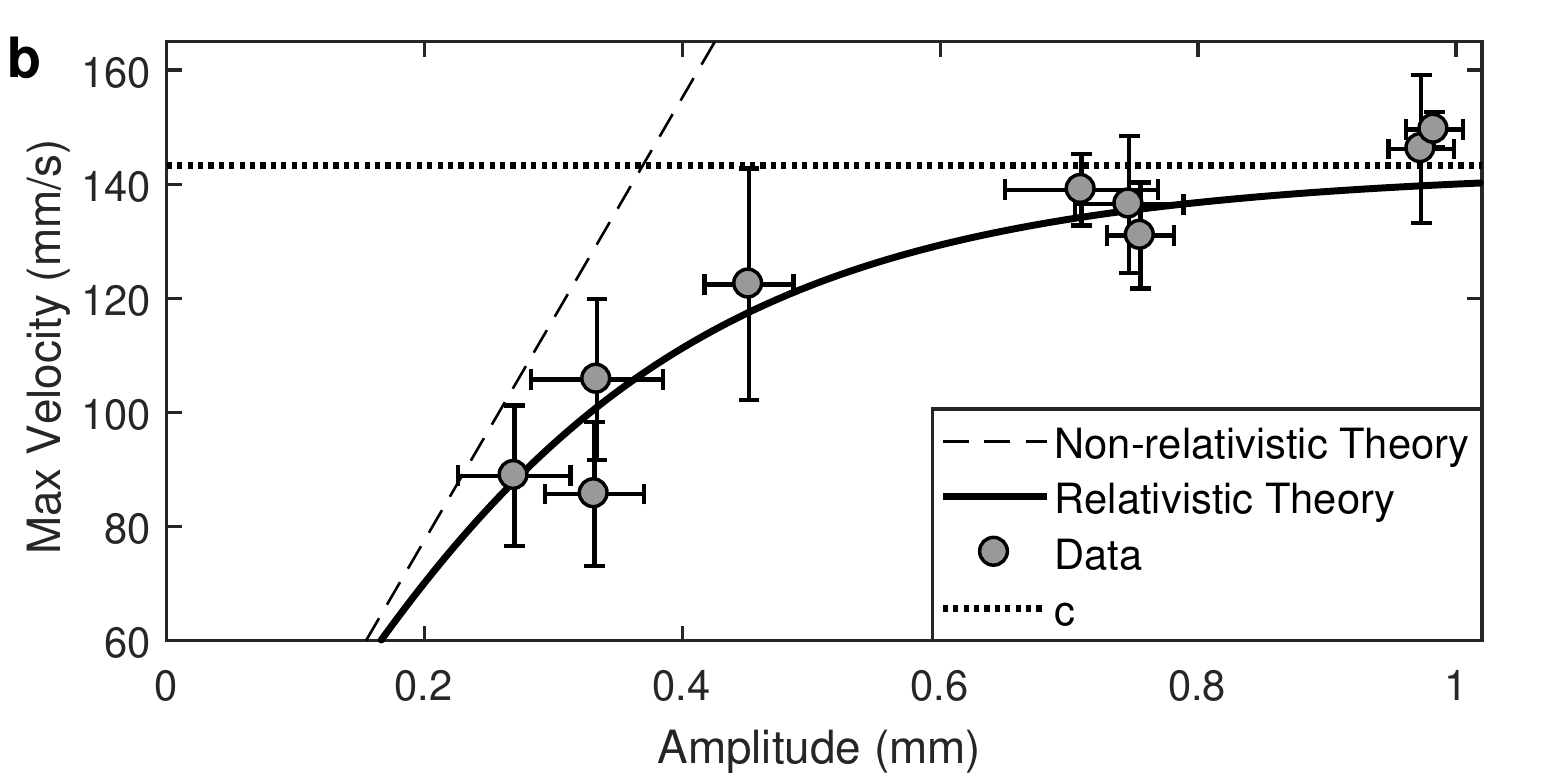}
\end{center}
\caption{
Measuring relativistic harmonic oscillator dynamics. \textbf{a:}~Comparison of measured and theoretical trajectories. Measured normalized mean atomic position versus time for $\beta_\mathrm{max} = 0.7$ (blue circles), 0.94 (red squares), and 0.97 (yellow triangles). Solid lines show theoretically predicted evolution according to Eqs.~\ref{eq:worldlines1}~and~\ref{eq:worldlines2}, with no adjustable fit parameters. 
\textbf{b:}~Amplitude-velocity relation of the relativistic harmonic oscillator. Points show measured peak velocity versus measured amplitude. Amplitude error bars indicate the $1/e$ ensemble radius and velocity error bars indicate estimated uncertainty of numerical differentiation. Lines show the non-relativistic expectation (dashed), the exact relativistic prediction (solid), and the effective speed of light (dotted). }
\label{fig:doubleworldlines}
\end{figure}

Beyond simply reproducing qualitative features of relativistic dynamics, this system allows a direct quantitative experimental test of the theory of the relativistic harmonic oscillator. Figure~\ref{fig:doubleworldlines}a shows the measured trajectories for oscillators at three maximum velocities corresponding to $\beta_\mathrm{max}=$ 0.7, 0.94, and 0.97. Predicted trajectories from equation~\ref{eq:worldlines1} are plotted as solid lines, without the use of any fitting parameter. The observed evolution of trajectory shape, from a sinusoidal and approximately harmonic form at low energies to an increasingly triangular form with a growing period, is in good quantitative agreement with the predictions of the relativistic theory.

A fundamental feature of the relativistic harmonic oscillator is that the maximum velocity saturates at the effective speed of light, with the amplitude diverging at finite maximum velocity rather than increasing linearly with maximum velocity as in the non-relativistic case. Figure~\ref{fig:doubleworldlines}b presents an experimental measurement of this phenomenon, showing the velocity asymptotically approaching $c$ as the amplitude increases. Results are consistent with the relativistic theory to within the estimated measurement error.  Here, velocity and amplitude were derived directly from measured position-versus-time datasets like those plotted in figure~\ref{fig:doubleworldlines}a; this procedure avoids the use of fits or theoretical assumptions, though it does give rise to sizable velocity error bars due to the noise-intolerance typically associated with numerical differentiation.

\begin{figure}[t!]
\begin{center}
\includegraphics[width=0.7\linewidth]{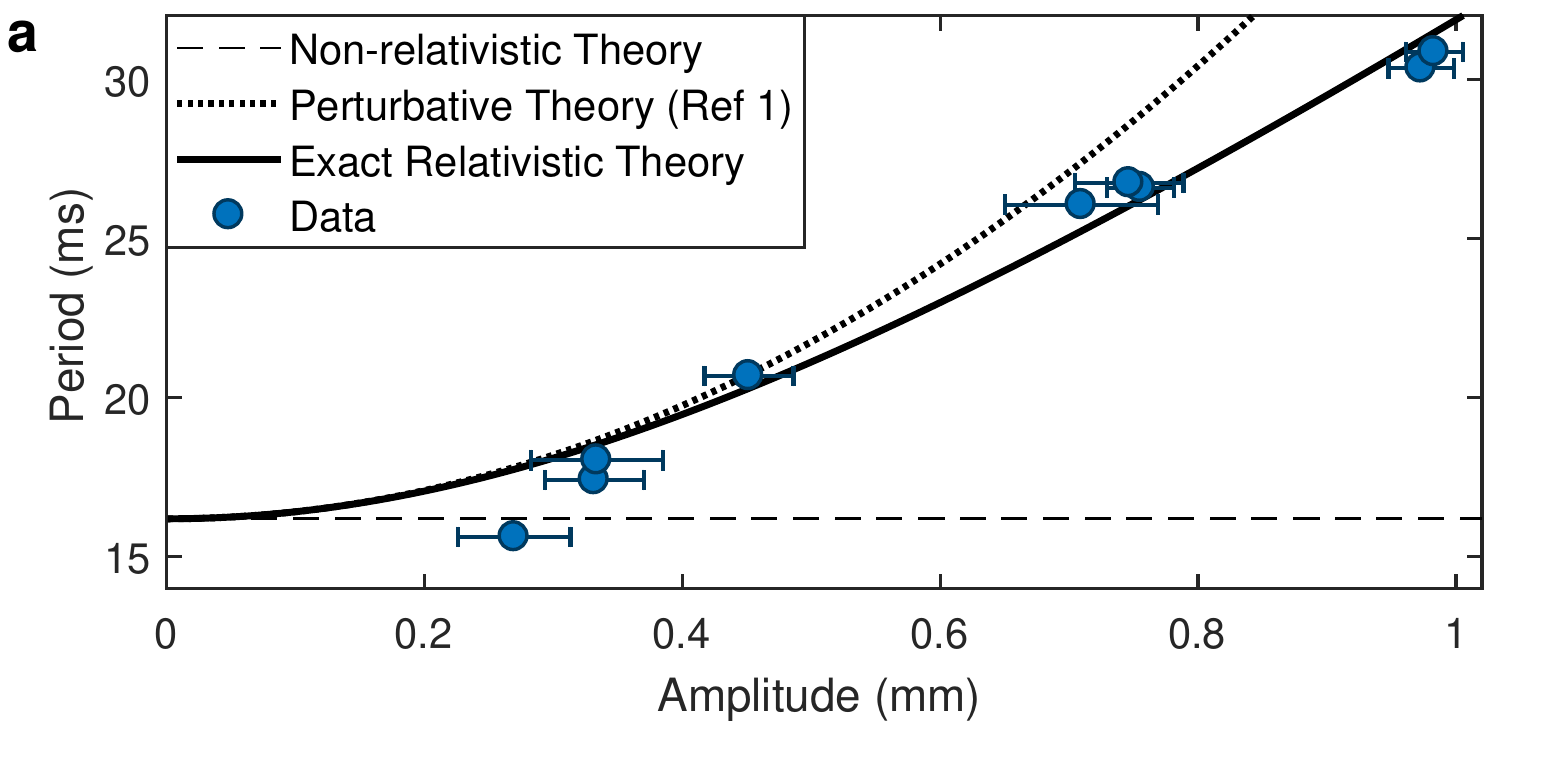}
\includegraphics[width=0.7\linewidth]{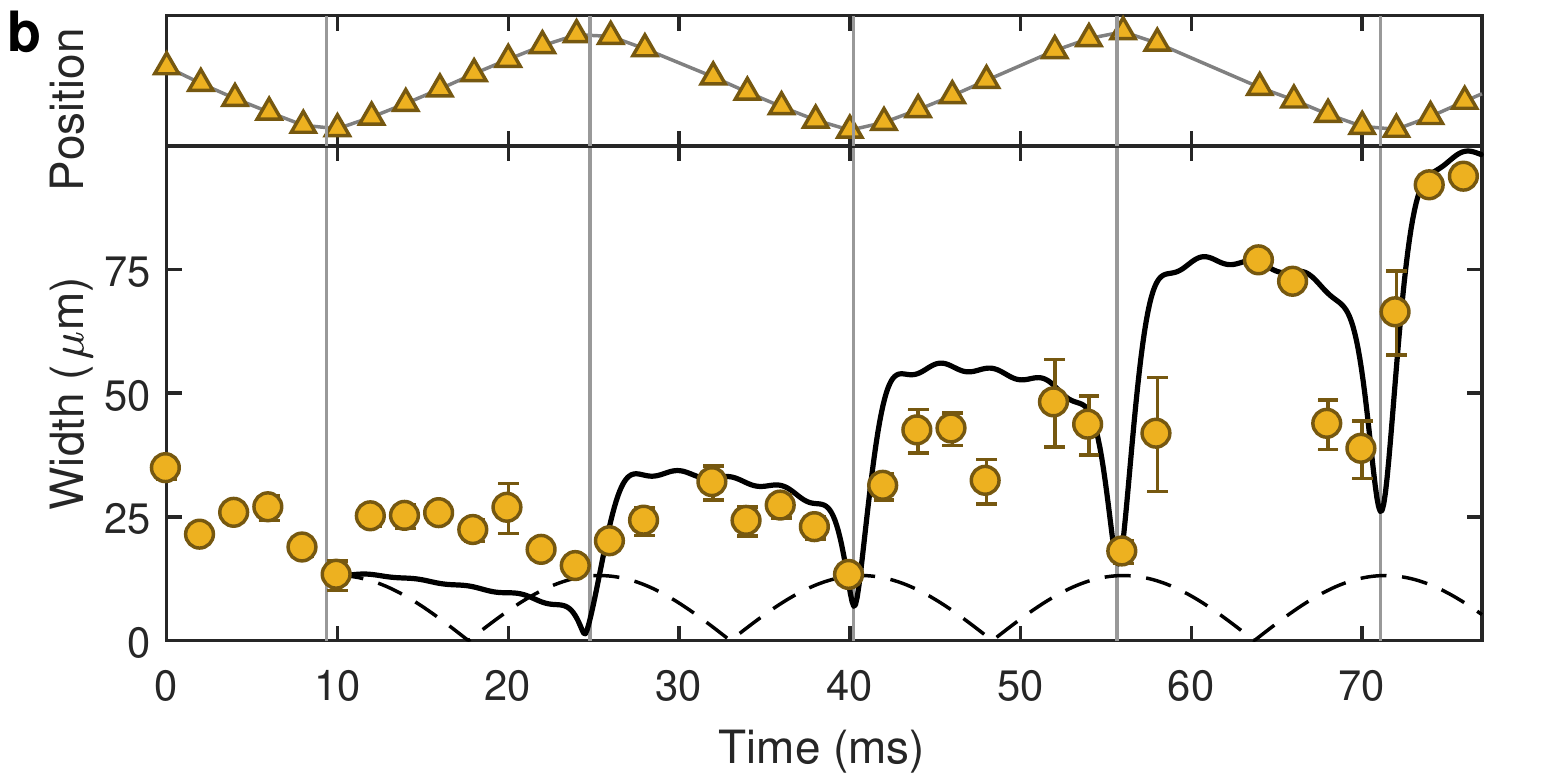}
\end{center}
\caption{
Measuring the effects of relativistic anharmonicity. \textbf{a:}~Measured oscillation period versus amplitude (circles). Amplitude error bars indicate the $1/e$ cloud radius. The non-relativistic harmonic theory (dashed), the leading-order theoretical prediction of reference~[1] and equation~\ref{eqperiod} (dotted) and the exact theoretical prediction derived from equation~\ref{eq:worldlines1} (solid) are plotted, without adjustable fit parameters. 
\textbf{b:}~Relativistic monopole oscillation and dephasing of an oscillator ensemble.  Top panel: center-of-mass dipole oscillations of an oscillator ensemble with average $\beta_{\mathrm{max}}=0.98$. Bottom panel: measured $1/e$ width versus time of the ensemble (circles). Error bars indicate estimated 95\% confidence interval.  The non-relativistic (dashed) and relativistic (solid) predictions for ensemble width evolution are plotted from the first turning point without adjustable fit parameters. 
}
\label{fig:doubledecoherence}
\end{figure}

Relativistic anharmonicity can also be quantitatively measured and compared to theory; we find that the experiment is a sufficiently accurate realization of a relativistic oscillator to distinguish beyond-leading-order corrections at high energy. Figure~\ref{fig:doubledecoherence}a shows the measured oscillation period $T$ as a function of measured oscillation amplitude $A$.  The leading-order prediction of reference~[1] and equation~\ref{eqperiod}, plotted as a dashed line, matches the data at low values of $A$ and $\gamma_0$ but predicts too high a period for the highest-energy oscillators. As the maximum Lorentz factor $\gamma_0$ increases, higher-order corrections become important. The exact prediction, calculated by setting $\phi=2\pi$ in equation~\ref{eq:worldlines1} and evaluating the elliptic integrals, is plotted as a solid line. Measured periods are in close agreement with the exact theoretical predictions for relativistic anharmonicity. 

We observe that relativistic effects modify the dynamics of oscillator ensembles initialized with a spread of energies: the ensembles dephase due to relativistic anharmonicity, and exhibit a monopole oscillation with phase  opposite to that of a non-relativistic oscillator ensemble. The initial atomic spatial distribution inherited from the Bose-Einstein condensate gives rise to a distribution of frequencies with a width that depends on the relativistic anharmonicity at the average amplitude. As the members of the oscillator ensemble acquire varying amounts of phase, the dephasing manifests first as an exponentially-enveloped increase in the width of the spatial distribution. A quantitative theoretical prediction for the time-dependent spatial width can be derived numerically by applying Eqs.~\ref{eq:worldlines1}~and~\ref{eq:worldlines2} to an appropriate initial ensemble, starting from the first turning point. Figure~\ref{fig:doubledecoherence}b shows a comparison between such a prediction and the measured time-dependent width for an oscillator ensemble with average energy deeply in the relativistic regime. The observed increase in width is quantitatively well-matched by the theory without any adjustable fit parameters, indicating that the dephasing of the oscillator ensemble is dominated by relativistic effects. In addition to the overall increase in width, both theory and experiment show oscillations of the width at twice the average oscillator frequency. These monopole oscillations are also a consequence of  relativistic anharmonicity; intriguingly, they have the opposite phase from non-relativistic monopole oscillations. The trajectories of two non-relativistic harmonic oscillators initialized at the same phase but slightly different energies will cross at the origin, leading to minimum ensemble width at the minimum of the potential. In contrast, the trajectories of two similarly initialized highly relativistic oscillators will cross near the turning points, leading to minimum ensemble width there. As figure~\ref{fig:doubledecoherence} demonstrates, the observed minimum ensemble width of the relativistic monopole oscillations indeed occurs at the turning points; this is exactly the opposite of the expected behavior for a non-relativistic oscillator.  

In conclusion, we have experimentally realized and quantitatively studied a harmonic oscillator in the relativistic regime using ultracold lithium atoms in the third band of an optical lattice. Though relativistic harmonic motion has been studied theoretically for decades, to the best of our knowledge this represents the first experimental observation of relativistic harmonic oscillator worldlines and dynamics. The measured worldline shapes, relativistic anharmonicity, monopole oscillations, and relativistic dephasing of oscillator ensembles are in good quantitative agreement with  relativistic predictions. 

\ack
The authors thank Peter Dotti, Sean Frazier, Vyacheslav Lebedev, Ethan Simmons, Yi Zeng, James Chow, Jacob Hines, and Andrew Ballin for experimental assistance, Gil Refael for useful discussion, and David Patterson for a critical reading of the manuscript, and acknowledge support from the Army Research Office (PECASE W911NF1410154 and MURI W911NF1710323), National Science Foundation (CAREER 1555313), and a President's Research Catalyst Award (CA-15-327861) from the UC Office of the President.

\vspace{.15in}

\bibliographystyle{iopart-num}
\providecommand{\newblock}{}

\end{document}